\documentclass{elsart}
\usepackage{graphics}
\begin{document}
\begin{frontmatter}

%%% article title
\title{Three--dimensional oscillator in magnetic field:
The de Haas--van Alphen effect in mesoscopic systems.}
\author[Radium]{N.K. Kuzmenko\corauthref{cor}},
\corauth[cor]{Corresponding author.}
\ead{Kuzmenko@NK9433.spb.edu}
\author[Univ]{V.M. Mikhajlov}
\address[Radium]{V.G.Khlopin Radium Institute, 194021
St.-Petersburg, Russia}
\address[Univ]{Institute of Physics St.--Petersburg State
University 198904, Russia}

%%% abstract
\begin{abstract}
The theoretical investigation of the cluster de Haas - van Alphen
(dHvA) oscillations in three-dimensional systems performed for
the first time. Applying a three-dimensional oscillator model to
systems with electron numbers $10<~N~\leq~10^5$ we predict
distinctive size effects: the dHvA oscillations can be observed
only within a certain temperature range determined by $N$; the
lower size limit for $N$ is $\approx 20$; the amount of the dHvA
oscillations is reduced with decreasing $N$ which is accompanied
by stretching the period of the oscillations.
\end{abstract}

\end{frontmatter}

\section{Introduction}
With increasing a uniform magnetic field applied to a macroscopic
body its electron characteristics can reveal low-temperature
oscillations manifesting themselves e.g. in the de Haas - van
Alphen (dHvA) or Schubnikov-de Haas effects \cite{shoenberg}.
Oscillations in magnetization of metal thin films, the period of
which increases with decreasing thickness, were  theoretically
predicted in Refs.~\cite{kosevitch,childers}. The Shubnikov-de
Haas oscillations were experimentally observed in antimony plates
(thickness is of order of $10^2\div 10^3$~$nm$) \cite{gajdukov}.
Now the dHvA effect is rather intensively studied in
two-dimensional mesoscopic systems (e.g.
\cite{yoshioka,bogachek,gudmundsson} and works cited therein).

However the possibility of the dHvA oscillations in atomic
clusters has not been established so far. In this Letter we
predict the properties of the dHvA oscillations in
three-dimensional ($3D$)-mesoscopic systems in the absence of the
disorder and interactions assuming that noninteracting electrons
are confined in an oscillator well. We find that in
$3D$-mesoscopic systems, where electrons confined in all three
directions, along with the dHvA oscillations there exist much
more types oscillations than in bulk metals, thin films and
two-dimensional mesoscopic systems. We consider the finite size
effects in the dHvA oscillations in clusters with electron
numbers $10~<~N~\leq~10^5$ in a wide range of arbitrarily oriented
magnetic fields and find that number of the dHvA oscillations is
reduced with decreasing $N$ and these oscillations appear at a
temperature which depends on $N$. From our calculations follows
that grains of materials with small the Fermi energy and the
effective electron mass are best suited to observe the cluster
dHvA oscillations.

\section{Model}
The harmonic oscillator model is widely exploited in describing
structure of quantal systems including mesoscopic ones
\cite{childers,felderhof,heiss,kouw}. This is caused by the
possibility of obtaining simple analytical solutions for
oscillator energy eigenvalues and wave functions. Electron motion
in the anysotropic $3D$~-~oscillator
potential and a uniform magnetic field $\vec{B}$
is described by the Hamiltonian

\begin{equation}
\label{Ham}
H=\sum_{\nu=x,y,z}\{\frac{1}{2m^*}(p_{\nu}-\frac{e}{c}A_{\nu})^2
+\frac{m^*}{2}\Omega_{\nu}^2r_{\nu}^2
+\frac{g}{2}\mu_B^*\sigma_{\nu}B_{\nu} \},
\end{equation}

($\vec{A}=\left[\vec{r}\times\vec{B}\right]/2$, $m^*$ is the
effective electron mass, $\Omega_{\nu}$ are oscillator
frequencies at $B=0$, $g$ is the effective Lande factor,
$\mu_B^*=e\hbar/2m^*c$, $\sigma_{\nu}$ is the Pauli matrix).
Eq.~(\ref{Ham}) being expressed through oscillator quanta,
$b^{\dag}_{\mu}$, $b_{\nu}$ ($\mu$, $\nu=x,y,z$) is a bilinear
form of these operators ($b^{\dag}b$, $b^{\dag}b^{\dag}$, $bb$).
We use the boson Bogolubov transformation to remove terms with two
creation and annihilation bosons and at the same time to
diagonalize the rest part, i.e. to reduce it to $\sum
W_{\alpha}b^{\dag}_{\alpha}b_{\alpha}$, $\alpha =1,2,3$ where
$W_{\alpha}$ are the eigen frequencies of new oscillator quanta
in arbitrarily oriented field. As known both fermion and boson
Bogolubov transformations result in equations quadratic in
eigenvalues. Therefore we arrive at the cubic equation in $W^2$
since Eq.~(\ref{Ham}) deals with three spatial
degrees of freedom. Relative simplicity of the
Hamiltonian~(\ref{Ham}) is embodied in a rather compact equation
for eigenvalues:
\begin{equation}
\label{eqcubic}
\prod_{\nu=x,y,z}(\Omega_{\nu}^2 - W^2)- W^2\sum_{\nu=x,y,z}
\omega_{\nu}^2(\Omega_{\nu}^2- W^2)=0,
\end{equation}
$\vec{\omega}$ is the cyclotron frequency,
$\vec{\omega}=2\mu_B^*\vec{B}$.

Three solutions of
Eq.~(\ref{eqcubic}), $W_{+}$, $W_{-}$, $W_0$, determine the
single electron energies (hereafter $\hbar=1$)
\begin{equation}
\label{eqeigen}
\varepsilon (n_{+},n_{-},n_{0},\omega)= W_+(n_+
+\frac{1}{2}) +  W_-(n_- +\frac{1}{2}) + W_0(n_0 +\frac{1}{2})
\pm\frac{g}{4}\omega.
\end{equation}
$n_{\pm}$, $n_0$ are integers.
The spin contributes to the
last term in Eq.~(\ref{eqeigen}).
Solutions $W_{\pm}$, $W_{0}$ vary with increasing $\omega$:
$W_{+}$  grows, $W_{-}$ falls down and $W_{0}$ depends mainly
on the direction of $\vec{B}$.
If the field is directed along one of the symmetry axes of the
spheroid, e.g. $\omega_z=\omega$, $\omega_x=\omega_y=0$,
$W_0=\Omega_z$ and $W_{\pm}$ are found straightforwardly. For
axial symmetry ($\Omega_x=\Omega_y$) the solutions coincide with
those given by Fock \cite{fock}.
In high fields ($\omega^2\gg\sum_{\nu}\Omega_{\nu}^2$) solutions
of Eq.~(\ref{eqcubic}) gain the values:
\begin{eqnarray}
\label{eqhigh}
W_{+}^2=\omega^2; \; \; \; \; \;
W_{-}^2=\frac{D}{\omega^2}; \; \; \; \; \;
W_{0}^2=\Omega_0^2; \\
\Omega_{0}^2=\sum_{\nu}\omega_{\nu}^2\Omega_{\nu}^2 /\omega^2;
\; \; \;
D=\prod_{\nu}\Omega_{\nu}^2/\Omega_{0}^2, \; \; \;
\nu=x,y,z.
\nonumber
\end{eqnarray}

To model single--electron levels
we adopt the relation \cite{bohr} between semiaxes
of ellipsoidal $3D$~-~clusters
($a_x$, $a_y$, $a_z$) and their oscillator frequencies
\begin{equation}
\label{eqomega}
\Omega_xa_x=\Omega_ya_y=\Omega_za_z; \;\;\;\;
\Omega=\frac{\varepsilon_F}{(3N)^{1/3}},
\end{equation}
$a_xa_ya_z=R^3$, $R=r_sN^{1/3}$, $r_s$ is the radius of a sphere
with the volume $V/N$.
$\Omega^3=\Omega_x\Omega_y\Omega_z=W_+W_-W_0$ is an invariant.
$\varepsilon_F$ is the Fermi energy. Eq.~(\ref{eqomega}) is true
for comparable semiaxes and small frequencies,
$\Omega_{x,y,z}\ll\varepsilon_F$.

\section{Results of calculations}
Our calculations of the electronic susceptibility ($\chi$) and
heat capacity ($C$) are performed for ellipsoidal clusters.
Values of $C$ are presented here in order to demonstrate the
temperature averaged level density near the Fermi level, which is
practically  proportional to $C$. Calculations involve so wide
single-electron spectra that their enlarging has no effect on
results presented below.

\begin{figure}[p]
\scalebox{0.5}{\includegraphics{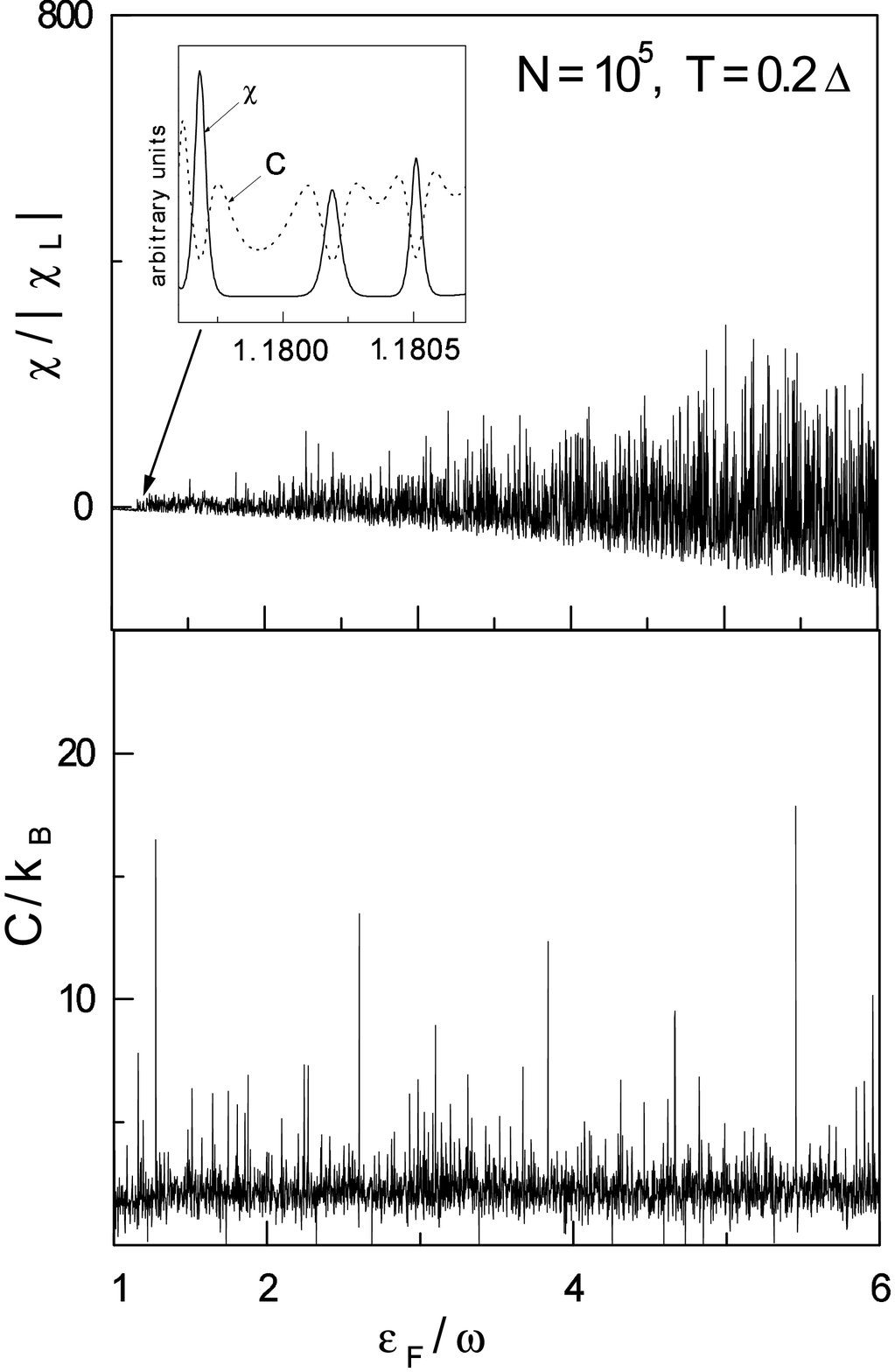}} \caption {
 Low temperature oscillations of the electron
susceptibility $\chi$ (a) and heat capacity $C$ (b) in a
tilted magnetic field ($\omega_z=\omega\cos\theta$,
$\omega_x=\omega\sin\theta\cos\varphi$,
$\omega_y=\omega\sin\theta\sin\varphi$, $\theta=\pi/4$,
$\varphi=\pi/4$) vs $\varepsilon_F/\omega$ for $N=10^5$ electrons
moving in an anysotropic oscillator potential ($a_y/a_x=1.33$,
$a_z/a_x=1.55$). $\chi_L~=~-\mu_B^2 N/2V\varepsilon_F$,
$\Delta=\varepsilon_F/3N$, $g=2$.}{\label{fig1}}
\end{figure}

\begin{figure}[p]
\scalebox{0.5}{\includegraphics{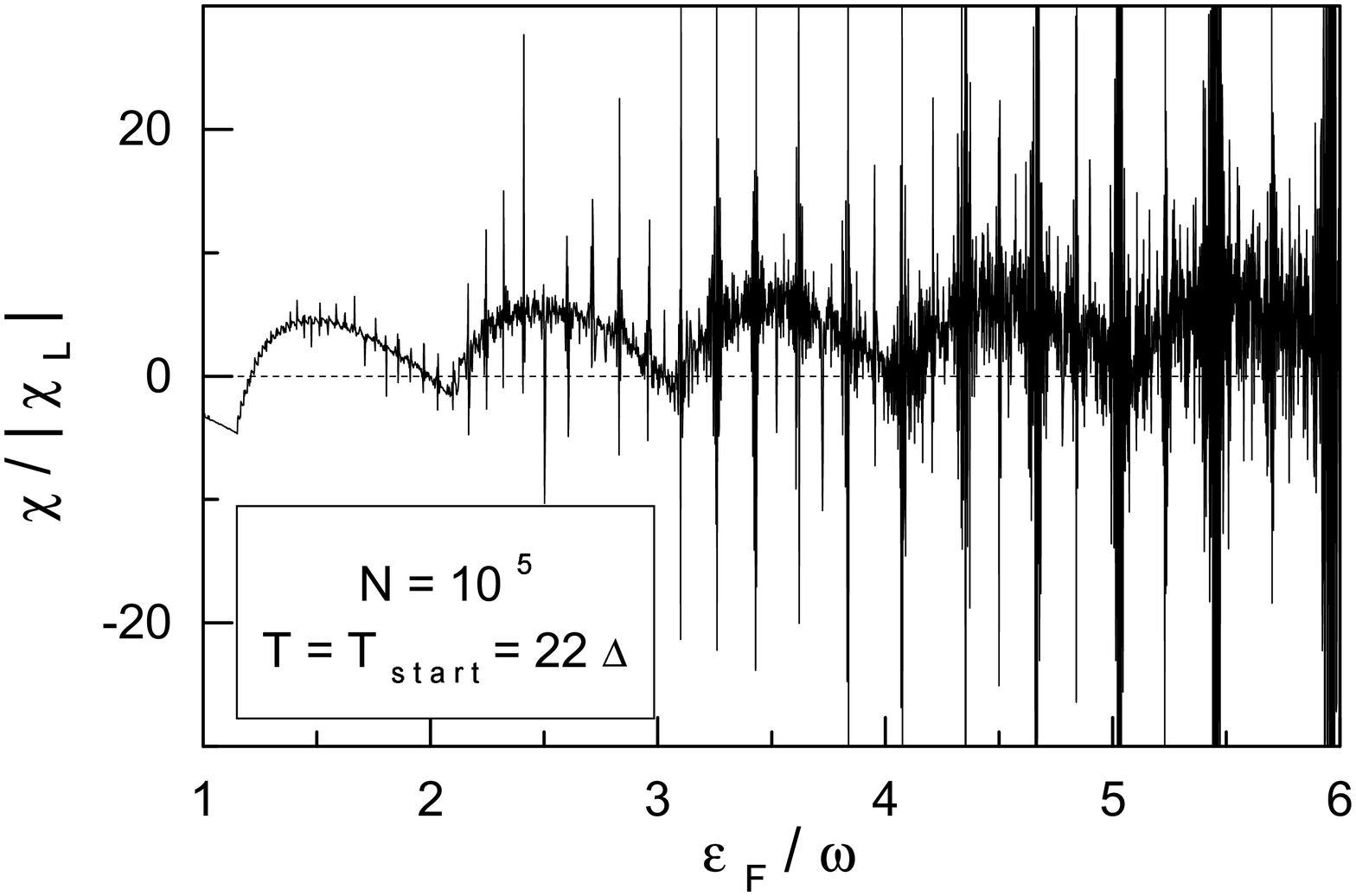}} \caption {
 Oscillations of the magnetic susceptibility
vs $\varepsilon_F/\omega$ for a cluster
with $N=10^5$ electrons at $T=T_{\mathrm{start}}=22\Delta$. The
oscillator anysotropy  and the direction of the magnetic field
are the same as in Fig.~\ref{fig1}, $g=2$.}{\label{fig2}}
\end{figure}

\begin{figure}[p]
\scalebox{0.5}{\includegraphics{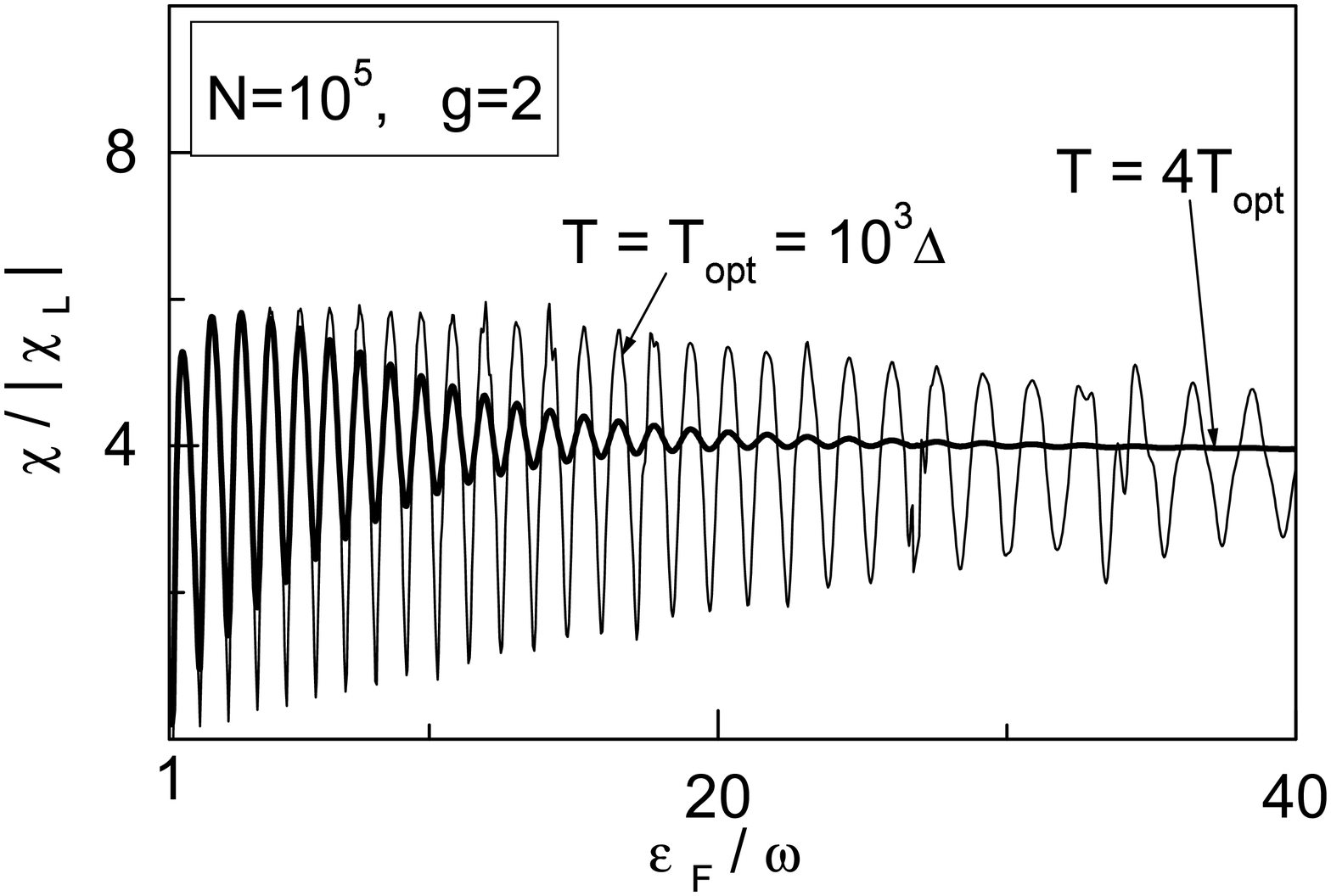}}
\caption {
 Influence of the temperature on the dHvA
oscillations in a system with $N=10^5$. The oscillator anysotropy
and the direction of the magnetic field are the same as in
Fig.~\ref{fig1}.}{\label{temperat}}
\end{figure}

\begin{figure}[p]
\scalebox{0.5}{\includegraphics{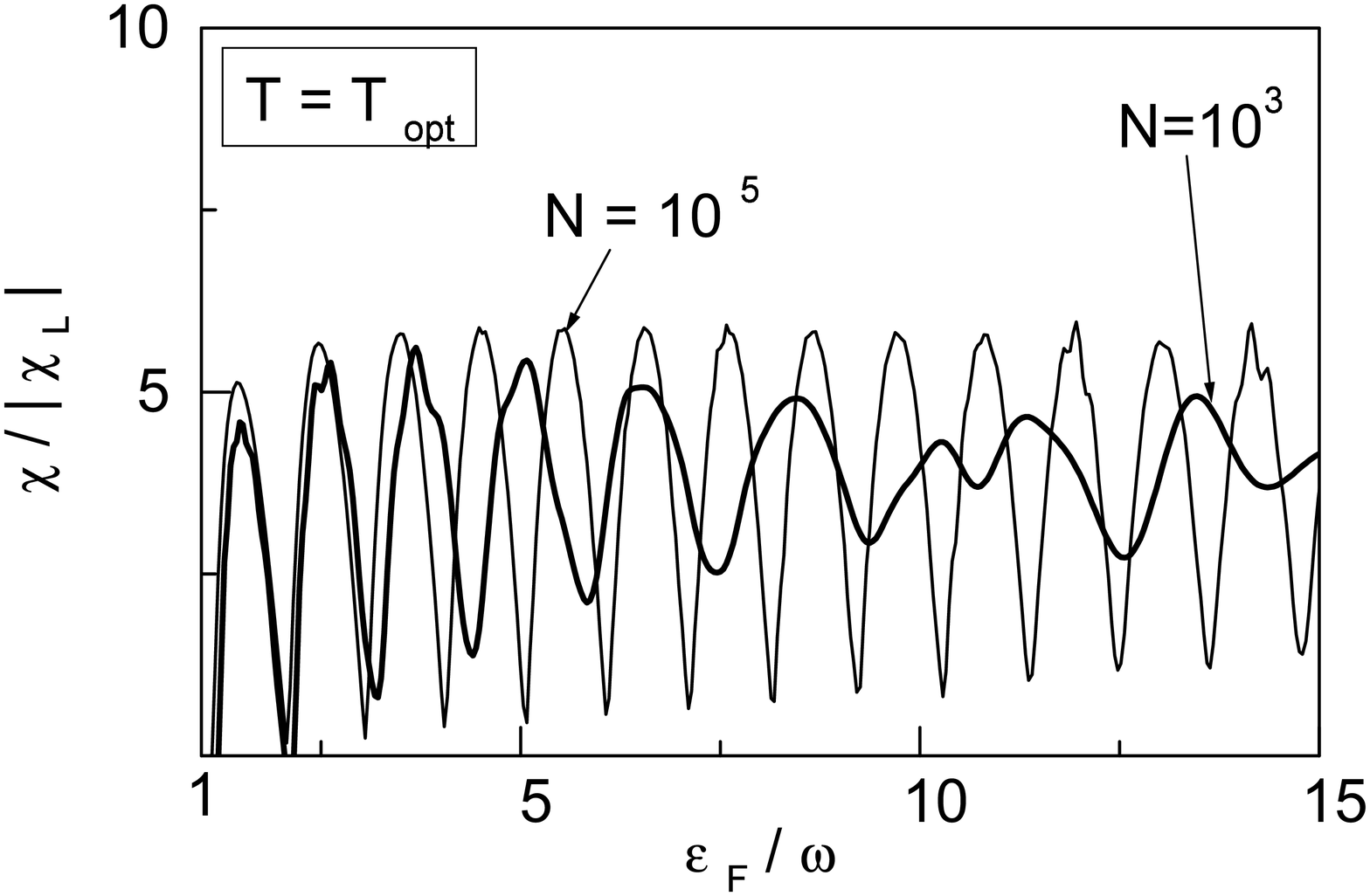}}
\caption {
{\label{comparis}} Comparison of the dHvA oscillations in systems
with $N=10^5$ and $N=10^3$ at $T=T_{\mathrm{opt}}$.
$T_{\mathrm{opt}}=10^3\Delta$ for $N=10^5$ and
$T_{\mathrm{opt}}=50\Delta$ for $N=10^3$. The oscillator
anysotropy and the direction of the magnetic field are the same
as in Fig.~\ref{fig1}, $g=2$.}
\end{figure}

\begin{figure}[p]
\scalebox{0.5}{\includegraphics{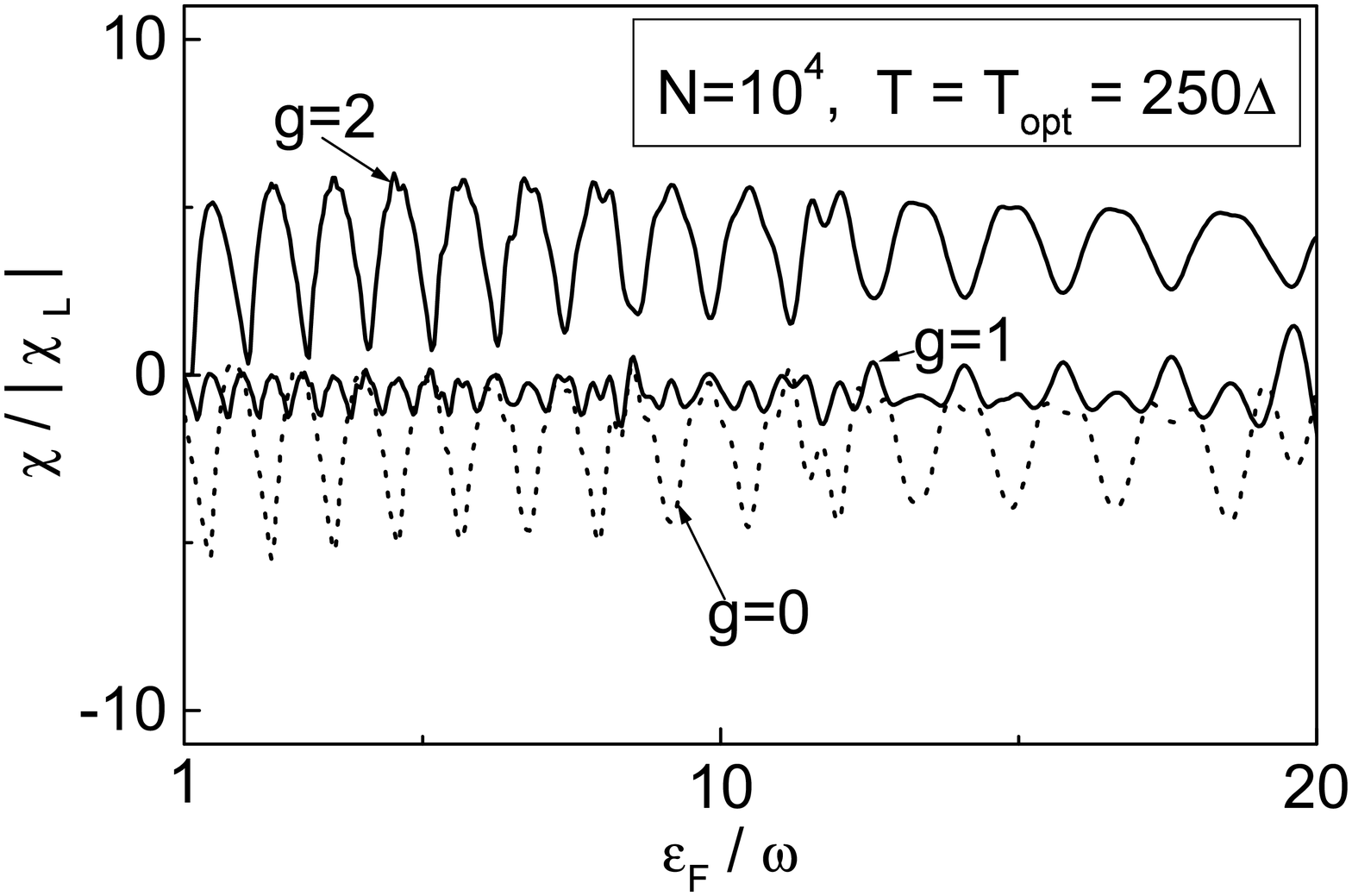}}
\caption {
 Influence of the effective Lande factor on the
dHvA oscillations in systems with $N=10^4$ at
$T=T_{\mathrm{opt}}=250\Delta$. The oscillator anysotropy  and
the direction of the magnetic field are the same as in
Fig.~\ref{fig1}.}{\label{gfactor}}
\end{figure}

\begin{figure}[p]
\scalebox{0.5}{\includegraphics{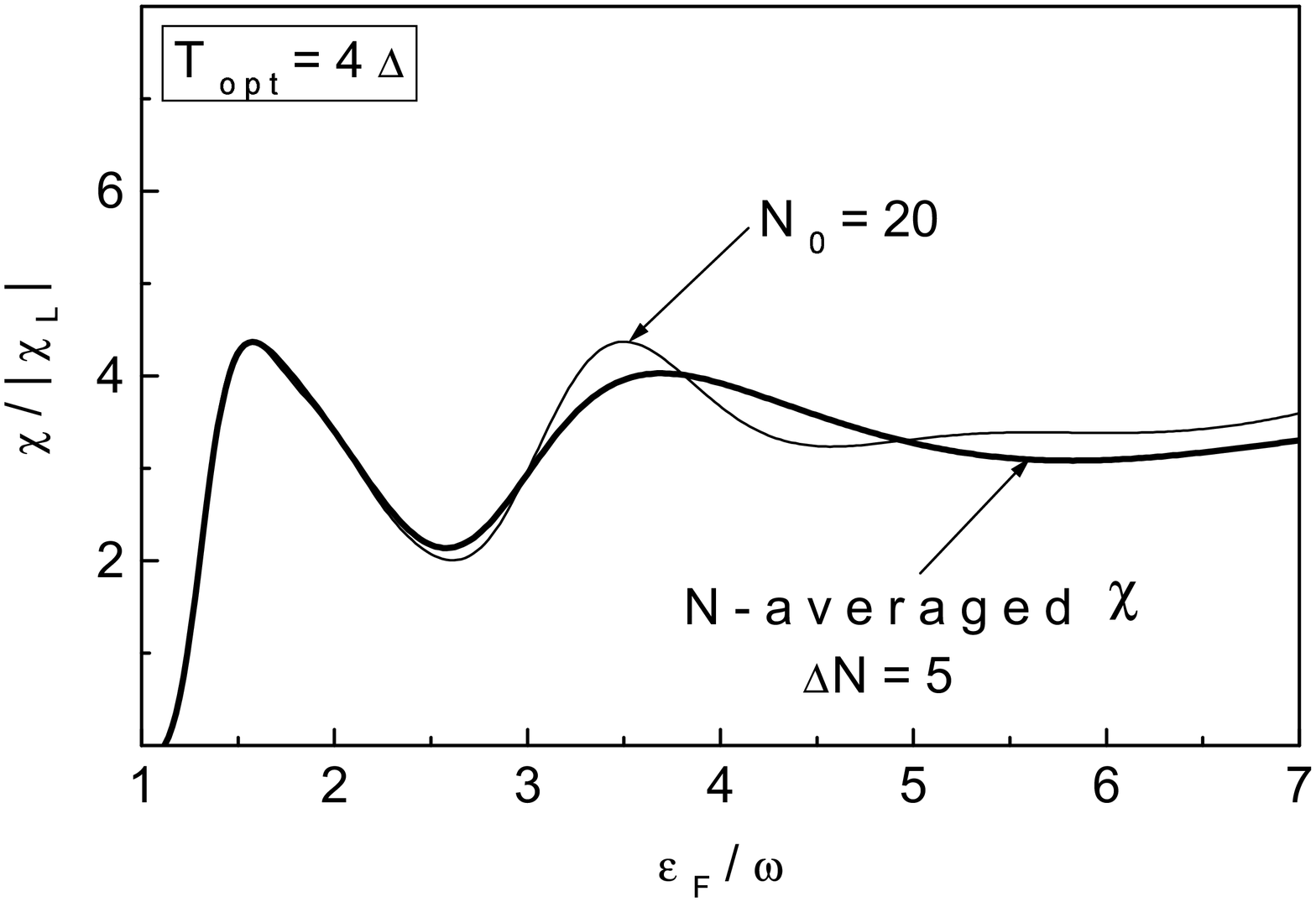}} \caption {
 N-averaged dHvA oscillations of spherical
clusters with $N=N_0\pm 3\Delta N$ ($N_0=20$, $\Delta N$ is the
averaging width parameter), $g=2$.}{\label{naveraged}}
\end{figure}

At very low temperatures ($T<\Delta$, $\Delta$ is a mean level
spacing) $\chi$ and $C$ are strongly oscillating functions of
$\varepsilon_F/\omega$ revealing no evident periodicity v.s. the
field (Fig.~\ref{fig1}).
Numerous peaks in $\chi$ and $C$ result from crossings of the
Fermi level and an upper level nearest to it that gives at low
temperatures a maximum in $\chi$ and simultaneously a minimum in
$C$~\cite{kuzmenko} as shown in the insert in Fig.~\ref{fig1}.
These low temperature oscillations, being absent in bulk metals
and thin films, are caused by the discreteness of the electronic
level spectrum. Increasing temperature drastically changes this
picture damping high frequency oscillations. Starting with a
temperature $T_{\mathrm{start}}$ one can observe the emerging of
regularities near the quantum limit, $\varepsilon_F/\omega\sim 1$
(Figs.~\ref{fig2}).
The final suppression of high frequency oscillations occurs at an
optimum temperature $T_{\mathrm{opt}}$
(Figs.~\ref{temperat}-\ref{naveraged}) exposing a maximum amount
($n_{\mathrm{max}}$) of periodic oscillations. They are analogous
with the dHvA oscillations in bulk metals and are determined by
the  common physical reason: the crossings of the Landau levels
with the Fermi surface in a strong magnetic field. Further
increasing the temperature ($T\gg T_{\mathrm{opt}}$) leads to
damping of the cluster dHvA oscillations (Fig.~\ref{temperat}).

To interpret these results we have found analytical expressions
for the $3D$ oscillator level density $\rho_0(\varepsilon)$ and
twice integrated level density $\rho_2(\varepsilon)$ which
determine the grand canonical $C$ and $\chi$ at a fixed particle
number $N$
\begin{eqnarray}
\label{eqC} C=k_B\beta^2\int d\varepsilon\rho_0(\varepsilon)
\left[ \left(\varepsilon-\lambda\right)^2-\left(\beta
\frac{\partial\lambda}{\partial\beta}\right)^2 \right]
\Phi (\varepsilon); \\
\label{eqHi}
\frac{\chi}{\mid\chi_L\mid}=\frac{8\varepsilon_F}{N}\beta\int d
\varepsilon \left[ \frac{\partial^2\rho_2(\varepsilon
)}{\partial\omega^2}- \rho_0(\varepsilon
)\left(\frac{\partial\lambda}{\partial\omega}\right)^2 \right]
\Phi
(\varepsilon ); \\
\Phi (\varepsilon )=\e^{\beta(\varepsilon-\lambda)}/\left[ 1
+\e^{\beta (\varepsilon -\lambda )}\right ]^2, \;\;\;
\beta=1/k_BT, \nonumber
\end{eqnarray}

$\chi_L=-\mu_B^2 N/2V\varepsilon_F$, $V$ being the volume of the
system. $\lambda$ is the chemical potential practically equal to
$\varepsilon_F$ and weakly oscillating with $\omega$ in a wide
region of $\omega$ up to $\omega\sim\varepsilon_F$ where
$\lambda$ markedly decreases.

\begin{equation}
\label{eqrho}
\rho_n(\varepsilon)=\frac{1}{16\pi(i)^{3+n}}\int_{\pounds} \,
\frac{dq}{q^n}\e^{i\varepsilon q} \left[
\sin\frac{W_+q}{2}\sin\frac{W_-q}{2}\sin\frac{W_0q}{2}
\right]^{-1}.
\end{equation}
The integration path ($\pounds$) envelops all poles including
$q=0$. The residue of $\rho_n(\varepsilon)$ in $q=0$ gives (after
integrating over $\varepsilon$  in Eqs.~(\ref{eqC}),~(\ref{eqHi}))
smooth functions of $\omega$ for $C$ and $\chi$ ($C_S$, $\chi_S$)
\begin{eqnarray}
\label{eqCsHis} C_S =
k_B^2\pi^2TN\left(\frac{\lambda}{\varepsilon_F}\right)\left[
1+\left(\frac{g\omega}{4\lambda}\right)^2-\left(\frac{\sum_{a=x,y,z}
\Omega_a^2 +\omega^2}
{12\lambda^2}\right)\right], \\
\chi_S/\mid\chi_L\mid = \left(-2+\frac{3}{2}g^2\right)
\left(\frac{\lambda}{\varepsilon_F}\right)^2 .
\end{eqnarray}

These expressions are true at $T\ll\varepsilon_F$. The first
negative term in $\chi_S/\mid\chi_L\mid$ is the ``3D-oscillator''
Landau diamagnetism and the second ($\sim g^2$) term is the
``3D-oscillator'' Pauli paramagnetism. Their absolute values are
two times more than those for free electrons, nevertheless their
ratio at $g=2$ is equal to the well known value
($\chi_L/\chi_P=-1/3$). $C_S$ and $\chi_S$ are those middle levels
on which all types of oscillations are superimposed. At high
enough temperatures (but $T/\varepsilon_F\ll 1$)  the most part
of oscillations are suppressed and $C$ and $\chi$ become equal to
$C_S$ and $\chi_S$. Fig.~\ref{temperat} shows that at
$4T_{\mathrm{opt}}$ the oscillations in the system with $N=10^5$
disappear in weak fields ($\varepsilon_F/\omega\geq 25$) and
$\chi$ becomes equal to $\chi_S$.

Oscillating components of $C$ and $\chi$ (which are determined by
residues in points $q\ne0$) can be divided into several
constituent parts. The large scale oscillation, which are denote
as the cluster dHvA oscillations by analogy with bulk metal
oscillations, are described by functions $\cos(2\pi\lambda
n_+/W_+)\cos(\pi g\omega n_+/2W_+)n_+^{-2}$. For the dHvA
oscillations frequencies $W_+$, $W_-$,  $W_0$ must not be
multiple, i.e. $W_+/W_-\neq n_+/n_-$ and $W_+/W_0\neq n_+/n_0$.
Since $\lambda\leq\varepsilon_F$ and $W_+\simeq\omega$ for
$\omega\gg\Omega$ the period ($t_+$) of the main ``tone'' of
these oscillations ($n_+=1$) on the scale $\varepsilon_F/\omega$
is almost equal to 1:
\begin{displaymath}
\cos(2\pi\lambda/W_+)\equiv\cos\left(2\pi\frac{1}{t_+}
\frac{\varepsilon_F}{\omega}\right), \;\;\;\;
t_+=\frac{\varepsilon_F}{\lambda}\frac{W_+}{\omega},
\end{displaymath}

though with decreasing $\omega$ the period $t_+$ is stretched, and
the smaller is the particle number the stronger is the stretching.
In Fig.~\ref{comparis} this stretching is quite evident for
$N~=~10^3$.
For instance in a simple case $\omega_z=\omega$,
$\omega_x=\omega_y=0$
\begin{displaymath}
W_+/\omega\simeq 0.5\left\{\left[ 4(3N)^{-2/3}
\left(\frac{\varepsilon_F}{\omega}\right)^2+1
\right]^{1/2}+1\right\}.
\end{displaymath}
The role of the effective electron Lande factor, entering into
$\cos(\pi g\omega n_+/2W_+)$, turns out to be similar to that in
the bulk dHvA oscillations \cite{shoenberg} (Fig.~\ref{gfactor}):
At $g=2$ the oscillations are paramagnetic while at $g=0$ they are
diamagnetic. At $g=1$ the main tone ($n_+=1$) disappears in
strong fields that leads to lowering amplitudes  and doubling
frequencies. Thus experimental studying of the cluster dHvA
oscillations could give information about cluster $g$-factors.
Functions in Fig.~\ref{fig1}-\ref{comparis} are calculated with
$g=2$.

The temperature damping factor $D_{\chi}(T,n_a)$ of the cluster
dHvA oscillations ($n_a=n_+$) and the other types of oscillations
considered below has for the susceptibility the same form as for
the bulk metal oscillations \cite{shoenberg}

\begin{equation}
\label{eqDHi} D_{\chi}(T,n_a)=\frac{x_a}{\sinh x_a}; \;\;\;
x_a=2\pi^2k_BTn_a/W_a,
\end{equation}

$D_{\chi}(T,n_+)$-factor causes attenuation of the dHvA amplitudes
with decreasing $\omega$ as seen e.g. in Fig.~\ref{temperat}. At
$T\ll\omega$ the temperature  damping of the dHvA oscillations is
inessential that allows summing over $n_+$ to be performed. It
leads to parabolic  segments in $\chi$ and $C$ which are well
illustrated in Fig.~\ref{fig2} by the example of $\chi$.

The temperature damping of the heat capacity oscillations is
governed by $D_C(T,n_a)$

\begin{equation}
\label{eqDC} D_{C}(T,n_a)=\frac{3x_a}{\sinh x_a}\left[ 1- 2(\coth
x_a)^2 +\frac{2}{x_a}\coth x_a\right],
\end{equation}

$x_a$ takes the same values as in Eq.~(\ref{eqDHi}). When
$x_a\rightarrow 0$ $D_{C}(T,n_a)\rightarrow 1$, however this
function decreases with $x_a$ much faster than $D_{\chi}(T,n_a)$.
Therefore with decreasing $\omega$ a stronger damping of the
oscillations in $C$ is observed as compared with $\chi$.
Besides, the amplitude of the dHvA oscillations
in $C$ is proportional to $W_+^2$ that additionally smoothes these
oscillations with increasing $\varepsilon_F/\omega$.

Another type of high frequency oscillations is connected with
$W_-$-frequency through the function $\cos(2\pi\lambda n_-/W_-)$.
(Here frequencies must not be multiple again.) Thus the period of
this oscillations rapidly falls down with increasing $\omega$.
Therefore the temperature damping becomes stronger at large
$\omega$. Hence one can find $T_{\mathrm{start}}$ supposing that
this temperature is such that
$D_{\chi}(T_{\mathrm{start}},n_-)\simeq 0.1$ i.e. this type of
oscillation is essentially suppressed at $T_{\mathrm{start}}$. As
$D_{\chi}=0.1$ at $x\simeq 4.5$ and $W_-=\Omega^2/\omega$  near
the quantum limit one gains:

\begin{equation}
T_{\mathrm{start}}\simeq\frac{1}{3}(3N)^{1/3}\Delta.
\end{equation}

In the $3D$-case there could be exist one more type of
oscillations: $\cos(2\pi\frac{\lambda}{W_0}n_0)$ ($W_0/W_+\neq
n_0/n_+$; $W_0/W_-\neq n_0/n_-$). However these oscillations can
really reveal themselves when the direction of $\vec{B}$ does not
coincide with any of oscillator symmetry axes and only at small
$\omega$ because near the quantum limit $W_0$ does not depend on
$\omega$ and these oscillations do not arise.

In all three cases considered above we marked in brackets that
$W_+$, $W_-$, $W_0$ should not be multiple to each other.
Nevertheless, if $\omega$ varies continuously a pair of
frequencies or even all three can become multiple. When $\omega$
reaches such values that
\begin{displaymath}
\frac{n_+}{W_+}=\frac{n_-}{W_-}\neq\frac{n_0}{W_0} \;\;\; or
\;\;\; \frac{n_0}{W_0}=\frac{n_-}{W_-}\neq\frac{n_+}{W_+}
\end{displaymath}
these types oscillations are suppressed at $T=T_{\mathrm{start}}$
and $\omega\leq\varepsilon_F$. The third type of such oscillations
$n_+/W_+=n_0/W_0\neq n_-/W_-$ appears in such points of $\omega$
where the ratio of $W_+$ and $W_0$ is equal to an integer, if
$n_0=1$, i.e. these points are divided by an interval
approximately equal to $W_0\sim\Omega$. Therefore this type of
oscillations can be damped by the temperature if
$D_{\chi}(T_{\mathrm{opt}},n_0)\simeq 0.1$. The parameter $x$ in
this case at $n_0=1$ is $2\pi^2 T_{\mathrm{opt}}/W_0=4.5$:

\begin{equation}
T_{\mathrm{opt}}\simeq\frac{1}{4}(3N)^{2/3}\Delta.
\end{equation}

The last type of oscillations $W_+/W_0/W_-=n_+/n_0/n_-$ is also
damped because the minimum value of $n_-$ is 1 and at
$\omega>2\Omega$ the frequency $W_-\ll W_0$, i.e. $n_0\gg 2$ and
$D_{\chi}\ll 0.1$. Amplitudes of all oscillations with multiple
frequencies increase with decreasing $\omega$ and at small
$\omega\leq 2\Omega$ the temperature damping is neutralized. This
is confirmed by our calculations indicating that in a wide range
of $N$ the cluster dHvA oscillations can be separated from others
at $T=T_{\mathrm{opt}}$ when the cyclotron frequency
$\omega_{\mathrm{min}}$ is about two times larger than $\Omega$
Eq.~(\ref{eqomega}):
\begin{equation}
\label{omegamin} \omega_{\mathrm{min}}\simeq 2\Omega, \;\;\;\;
n_{\mathrm{max}}\simeq\frac{1}{2}(3N)^{1/3}.
\end{equation}
Eq.~(\ref{omegamin}) implies that these oscillations begin at such
magnetic field, the cyclotron radius $R_c$ of which turns out to
be equal to the minimal size of a cluster in the plane
perpendicular to the magnetic field. Thus $\omega_{\mathrm{min}}$
is determined by the same conditions as in bulk metals. The
existence of $n_{\mathrm{max}}$ depending on $N$ limits the
cluster particle number $N_{\mathrm{min}}$ at which the dHvA
oscillations might be still observed. Since the minimum value of
$n_{\mathrm{max}}$ is 1 (the only parabolic segment shows up in
$\chi$) Eq.~(\ref{omegamin}) gives $N_{\mathrm{min}}\simeq 20$
(Fig.~\ref{naveraged}).
Fig.~\ref{naveraged} and our calculations for $N\sim 10^4$ show
that for an array of clusters with wide enough size distribution (
and consequently with wide distribution in particle numbers) one
can observe practically the same number of the oscillations as for
a single cluster with $N_0$.

We have analyzed the dependence of our results on variations of
cluster shapes and the direction of the magnetic field. For this
purpose at a fixed particle number we have altered the cluster
shape from oblate to prolate ($0.1<a_z/a_x<3$) with different
degree of nonaxiality ($1<a_y/a_x<2$) and varied the direction of
a magnetic field.
Our calculations have shown that the properties of the dHvA
oscillations in $3D$ finite systems are rather insensitive to the
variations of the cluster shape if the size of a cluster in the
direction of the field is not too small as against the size in
the perpendicular plane (e.g. $a_z$ should be greater than
$0.5a_x$).
Thus
reasonably large variations of cluster shape and direction of the
magnetic field cannot hinder measurements of the cluster dHvA
oscillations.

Values of minimal magnetic fields required to observe cluster
dHvA oscillations at $T_{\mathrm{opt}}$ can be assessed from
Eq.~(\ref{omegamin}) (obviously the upper limit is
$\varepsilon_F$)
\begin{equation}
\omega_{\mathrm{min}}\sim \frac{\varepsilon_F}{2(3N)^{1/3}};
\;\;\;\; B_{\mathrm{min}}(T)\sim 10^4\frac{m^*}{m}
\frac{\varepsilon_F(eV)}{2(3N)^{1/3}}.\label{eq7}
\end{equation}
These equations indicate that reasonable fields can be used for
large grains of such materials for which the Fermi energy
($\varepsilon_F$) and effective electron mass ($m^*$) are rather
small quantities. In addition, at a small $\varepsilon_F$ and
large $N$ the temperatures can be low enough,
$T_{\mathrm{opt}}\sim 1K$, to neglect electron scattering effects.
For GaAs systems (m$^*=0.067m$, $\varepsilon_F\approx 10$ $meV$)

\begin{equation}
B_{\mathrm{min}}(T)\sim 2/N^{1/3}; \;\;\;
T_{\mathrm{opt}}(K)\sim
20/N^{1/3}.
\end{equation}

\section{Summary}
In conclusion using the three-dimensional oscillator model we
predict the dHvA oscillations in grains of materials with small
$\varepsilon_F$ and $m^*$. Analogously with bulk metals the
cluster dHvA oscillations arise when the cyclotron radius is less
than the size of a system. Size effects are clearly observed for
$N\leq 10^5$: the number of the oscillations is reduced with
decreasing $N$; the size limit for $N$ is $\sim 20$; the period
of the oscillations stretches with increasing
$\varepsilon_F/\omega$.
The properties of the dHvA oscillations (the number of
oscillations, the amplitude, the period) are weakly sensitive to
the significant variations of cluster shapes and the direction of
the magnetic field especially near the quantum limit.
The characteristic feature of the cluster dHvA oscillations is a
special temperature regime at which high frequency oscillations
caused by the discreteness of electronic level spectra are
suppressed. The oscillator model predicts that the
$3D$-mesoscopic dHvA oscillations appear at
$T_{\mathrm{start}}\simeq\varepsilon_F/3(3N)^{2/3}$ and the most
favorable temperature for the observation of their maximum number
is $T_{\mathrm{opt}}\simeq\varepsilon_F/4(3N)^{1/3}$, while an
excess of $T_{\mathrm{opt}}$ results in smoothing and subsequent
disappearing the cluster dHvA oscillations.

We thank V.E. Bunakov for useful discussions.

\end{document}